\documentclass[a4paper,12pt,showkeys]{revtex4}
\usepackage[utf8]{inputenc}
\usepackage[T1]{fontenc}
\usepackage{lmodern}
\usepackage{epsfig}
\usepackage{array}
\usepackage{amsmath}
\usepackage{epstopdf}
\usepackage{graphicx}
\graphicspath{/}
\begin{document}

\title{Prediction of neutron electromagnetic form factors behaviors
       just by the proton electromagnetic form factors data}
\date{\today}

\medskip

\author{Anna Z. Dubni\v ckov\'a}
\address{Department of Theoretical Physics, Comenius University,
Bratislava, Slovak Republic}

\author{Stanislav Dubni\v cka}
\address{Institute of Physics, Slovak Academy of Sciences,
Bratislava, Slovak Republic}

\begin{abstract}
   It is demonstrated how the neutron electromagnetic form factors behaviors in the whole region of their definition can be predicted
theoretically just by using the proton form factors data to be described by the advanced proton electromagnetic structure Unitary and Analytic model.
\end{abstract}


\maketitle

\section{Introduction}

   Dominant building particles of the whole universe are protons $p$ and electrons $e$. While the electron
is always stable and a point-like particle up to now, the proton, though also to be stable, is compound of three quarks and as a result in interaction with electrons
it appears as an extended object in the space. The latter property of the proton is called the proton electromagnetic (EM) structure, for the first time revealed in
the middle of the last century, before the quark model of strongly interacting particles, hadrons, has been established. Later on this
non-point-like nature of the proton has been generalized also to all other hadrons.

   The EM structure of any hadron is completely described by scalar functions to be called EM form factors (FFs) of one variable,
in the spacelike region it is the momentum transfer squared $t=-Q^2$ and in timelike region it is the total c.m. energy squared  $s=w^2$.

   The number of EM FFs depends on the spin of the hadron under consideration. The EM structure of a hadron with spin 0 is completely described
by one EM FF, the EM structure of a hadron with spin 1/2 is completely described by two EM FFs, the EM structure of a hadron with spin 1 is
completely described by three EM FFs. In the case of the EM structure of hadrons with spin higher than one a situation is even more complicated and generally it is not solved satisfactorily up to now.

   Further our interest will be concentrated to the $1/2^+$ octet baryons consisting of
the following particles: $p, n, \Lambda^0, \Sigma^+, \Sigma^0, \Sigma^-, \Xi^0$ and $\Xi^-$.

   To know their EM structure means to have adequate multitude of as precise as possible experimental information on their EM FFs in the
whole region of their definition and to dispose with reliable theoretical models to be able predict behaviors of these EM FFs consistently with data.

   However, from all members of the $1/2^+$ octet baryons the most reliable and dense experimental data exist on the proton EM structure just due to its stability.
They have been obtained by the processes like $e^-p \to e^-p$, $e^+p \to e^+p$, $e^+e^- \to p \bar p$, $p \bar p \to e^+e^-$ and polarizations there. And thanks to a progress with more accomplished technologies one can expect even improved situation with these data in near future.
Moreover the advanced proton EM structure $Unitary\&Analytic$ model \cite{ABDD} has been constructed recently, which reproduces all existing data quite well.

   The next member of the $1/2^+$ octet baryons is the neutron $n$, for which there are no very precise and reliable experimental data just due to
its instability. One is unable to prepare neither neutron targets nor antineutron beams for realization of similar experiments as they have been carried out with protons. In obtaining of the neutron EM FFs data a lot of model assumptions have been utilized and therefore finally the neutron data are not very reliable.

   Heaving in mind the latter, in this paper we would like to demonstrate how the neutron EM FFs behaviors are predicted
theoretically in the whole region of their definition by using the reliable proton EM FFs data, for a description of which the advanced proton EM structure $Unitary\&Analytic$ model \cite{ABDD} is applied.

   In the next papers a similar approach is planned to be realized also for other members of the $1/2^+$ octet baryons, so-called hyperons, starting with $\Lambda^0$ and
ending with $\Xi$ hyperons, however for numerical estimation of free physically interpretable parameters of their $Unitary\&Analytic$ models a completely different and more laborious approach is developed.

\section{Proton electromagnetic structure model}

   The EM structure of the proton is completely described theoretically by two
independent functions, the proton Dirac $F^p_1(t)$ and Pauli $F^p_2(t)$ FFs,
which naturally appear in a decomposition of the proton matrix element of the EM current $J^{EM}_\mu (0)$
in the form of coefficients of two linearly independent covariants constructed from the four momenta $p, p'$, $\gamma$-matrices and Dirac bi-spinors
\begin{small}
\begin{eqnarray}
  <p|J^{EM}_\mu (0)|p>=e \bar u(p')\{\gamma_\mu F^p_1(t)+\frac{i}{2m_p}
  \sigma_{\mu \nu}(p'-p)_\mu F^p_2(t)\} u(p),
\end{eqnarray}
\end{small}
with $m_p$ to be the proton mass.

    Description of the proton is even improved if the following mixed transformation properties of the EM current
$J^{EM}_\mu (0)$ under the rotation in the isospin space are utilized. One its part transforms like isoscalar and another like the third component of isovector.  These transformation properties lead to a splitting of the proton Dirac and Pauli EM FFs to flavour-independent isoscalar $F^N_{1s}(t), F^N_{2s}(t)$ and isovector $F^N_{1v}(t), F^N_{2v}(t)$ parts  as follows
\begin{small}
\begin{eqnarray}
 F^p_1(t)=[F^N_{1s}(t)+F^N_{1v}(t)]\nonumber\\
 F^p_2(t)=[F^N_{2s}(t)+F^N_{2v}(t)]\\\nonumber
\end{eqnarray}
\end{small}
whereby the sign between them is specified by the sign of the third component of the proton isospin to be $+1/2$.

   The FFs $F^N_{1s}(t), F^N_{1v}(t), F^N_{2s}(t), F^N_{2v}(t)$ are analytic in the whole complex t-plane
besides the cuts on the positive real axis starting for isovector FFs at the two-pion threshold and for isoscalar FFs at three-pion threshold.

   In the paper \cite{ABDD} the advanced 9 vector-meson resonance\\
 $\rho(770), \omega(782), \phi(1020); \rho'(1450), \omega'(1420), \phi'(1680); \rho''(1700), \omega''(1650), \phi''(2170);$  \cite{PDG}\\
Unitary and Analytic $(U\&A)$ model for proton isoscalar and isovector Dirac and Pauli FFs has been constructed
\begin{eqnarray}\label{FN1s}\nonumber
  F^N_{1s}[V(t)]=\Bigg(\frac{1-V^2}{1-V^2_N}\Bigg)^4\Bigg\{\frac{1}{2}H_{\omega''}(V)H_{\phi''}(V)\\\nonumber
  +\Bigg[H_{\phi''}(V)H_{\omega'}(V)\frac{(C^{1s}_{\phi''}-C^{1s}_{\omega'})}{(C^{1s}_{\phi''}-C^{1s}_{\omega''})}+
  H_{\omega''}(V)H_{\omega'}(V)\frac{(C^{1s}_{\omega''}-C^{1s}_{\omega'})}{(C^{1s}_{\omega''}-C^{1s}_{\phi''})}\\
  -H_{\omega''}(V)H_{\phi''}(V)\Bigg](f^{(1)}_{\omega'NN}/f_{\omega'})\nonumber\\
  +\Bigg[H_{\phi''}(V)H_{\phi'}(V)\frac{(C^{1s}_{\phi''}-C^{1s}_{\phi'})}{(C^{1s}_{\phi''}-C^{1s}_{\omega''})}+
  H_{\omega''}(V)H_{\phi'}(V)\frac{(C^{1s}_{\omega''}-C^{1s}_{\phi'})}{(C^{1s}_{\omega''}-C^{1s}_{\phi''})}\nonumber\\
  -H_{\omega''}(V)H_{\phi''}(V)\Bigg](f^{(1)}_{\phi'NN}/f_{\phi'})\\\nonumber
  +\Bigg[H_{\phi''}(V)L_{\omega}(V)\frac{(C^{1s}_{\phi''}-C^{1s}_{\omega})}{(C^{1s}_{\phi''}-C^{1s}_{\omega''})}+
  H_{\omega''}(V)L_{\omega}(V)\frac{(C^{1s}_{\omega''}-C^{1s}_{\omega})}{(C^{1s}_{\omega''}-C^{1s}_{\phi''})}\\\nonumber
  -H_{\omega''}(V)H_{\phi''}(V)\Bigg](f^{(1)}_{\omega NN}/f_{\omega})\\\nonumber
  +\Bigg[H_{\phi''}(V)L_{\phi}(V)\frac{(C^{1s}_{\phi''}-C^{1s}_{\phi})}{(C^{1s}_{\phi''}-C^{1s}_{\omega''})}+
  H_{\omega''}(V)L_{\phi}(V)\frac{(C^{1s}_{\omega''}-C^{1s}_{\phi})}{(C^{1s}_{\omega''}-C^{1s}_{\phi''})}\\
  -H_{\omega''}(V)H_{\phi''}(V)\Bigg](f^{(1)}_{\phi NN}/f_{\phi})\Bigg\}\nonumber
\end{eqnarray}
with 5 free parameters
$(f^{(1)}_{\omega'NN}/f_{\omega'}), (f^{(1)}_{\phi'NN}/f_{\phi'}),
(f^{(1)}_{\omega NN}/f_{\omega}), (f^{(1)}_{\phi NN}/f_{\phi}),
t^{1s}_{in}$
\begin{eqnarray}\label{FN1v}\nonumber
  F^N_{1v}[W(t)]=\Bigg(\frac{1-W^2}{1-W^2_N}\Bigg)^4\Bigg\{\frac{1}{2}L_\rho(W)L_{\rho'}(W)\\
  +\Bigg[L_{\rho'}(W)L_{\rho''}(W)\frac{(C^{1v}_{\rho'}-C^{1v}_{\rho''})}{(C^{1v}_{\rho'}-C^{1v}_\rho)}+
  L_\rho(W)L_{\rho''}(W)\frac{(C^{1v}_\rho-C^{1v}_{\rho''})}{(C^{1v}_\rho-C^{1v}_{\rho'})}\\\nonumber
  -L_\rho(W)L_{\rho'}(W)\Bigg](f^{(1)}_{\rho NN}/f_{\rho})\Bigg\}
\end{eqnarray}
with 2 free parameters
$(f^{(1)}_{\rho NN}/f_{\rho})$ and $t^{1v}_{in}$,
\begin{eqnarray}\label{FN2s}\nonumber
  F^N_{2s}[U(t)]=\Bigg(\frac{1-U^2}{1-U^2_N}\Bigg)^6\Bigg\{\frac{1}{2}(\mu_p+\mu_n-1)H_{\omega''}(U)H_{\phi''}(U)H_{\omega'}(U)\\\nonumber
  +\Bigg[H_{\phi''}(U)H_{\omega'}(U)H_{\phi'}(U)\frac{(C^{2s}_{\phi''}-C^{2s}_{\phi'})(C^{2s}_{\omega'}-C^{2s}_{\phi'})}
  {(C^{2s}_{\phi''}-C^{2s}_{\omega''})(C^{2s}_{\omega'}-C^{2s}_{\omega''})}\\\nonumber
  +H_{\omega''}(U)H_{\omega'}(U)H_{\phi'}(U)\frac{(C^{2s}_{\omega''}-C^{2s}_{\phi'})(C^{2s}_{\omega'}-C^{2s}_{\phi'})}
  {(C^{2s}_{\omega''}-C^{2s}_{\phi''})(C^{2s}_{\omega'}-C^{2s}_{\phi''})}\\\nonumber
  +H_{\omega''}(U)H_{\phi''}(U)H_{\phi'}(U)\frac{(C^{2s}_{\omega''}-C^{2s}_{\phi'})(C^{2s}_{\phi''}-C^{2s}_{\phi'})}
  {(C^{2s}_{\omega''}-C^{2s}_{\omega'})(C^{2s}_{\phi''}-C^{2s}_{\omega'})}\\\nonumber
  -H_{\omega''}(U)H_{\phi''}(U)H_{\omega'}(U)\Bigg](f^{(2)}_{\phi'NN}/f_{\phi'})\\\nonumber
  +\Bigg[H_{\phi''}(U)H_{\omega'}(U)L_{\omega}(U)\frac{(C^{2s}_{\phi''}-C^{2s}_{\omega})(C^{2s}_{\omega'}-C^{2s}_{\omega})}
  {(C^{2s}_{\phi''}-C^{2s}_{\omega''})(C^{2s}_{\omega'}-C^{2s}_{\omega''})}\\\nonumber
  +H_{\omega''}(U)H_{\omega'}(U)L_{\omega}(U)\frac{(C^{2s}_{\omega''}-C^{2s}_{\omega})(C^{2s}_{\omega'}-C^{2s}_{\omega})}
  {(C^{2s}_{\omega''}-C^{2s}_{\phi''})(C^{2s}_{\omega'}-C^{2s}_{\phi''})}+\\
  +H_{\omega''}(U)H_{\phi''}(U)L_{\omega}(U)\frac{(C^{2s}_{\omega''}-C^{2s}_{\omega})(C^{2s}_{\phi'}-C^{2s}_{\omega})}
  {(C^{2s}_{\omega''}-C^{2s}_{\omega'})(C^{2s}_{\phi''}-C^{2s}_{\omega'})}\\\nonumber
  -H_{\omega''}(U)H_{\phi''}(U)H_{\omega'}(U)\Bigg](f^{(2)}_{\omega NN}/f_{\omega})\\\nonumber
  +\Bigg[H_{\phi''}(U)H_{\omega'}(U)L_{\phi}(U)\frac{(C^{2s}_{\phi''}-C^{2s}_{\phi})(C^{2s}_{\omega'}-C^{2s}_{\phi})}
  {(C^{2s}_{\phi''}-C^{2s}_{\omega''})(C^{2s}_{\omega'}-C^{2s}_{\omega''})}\\\nonumber
  +H_{\omega''}(U)H_{\omega'}(U)L_{\phi}(U)\frac{(C^{2s}_{\omega''}-C^{2s}_{\phi})(C^{2s}_{\omega'}-C^{2s}_{\phi})}
  {(C^{2s}_{\omega''}-C^{2s}_{\phi''})(C^{2s}_{\omega'}-C^{2s}_{\phi''})}\\\nonumber
  +H_{\omega''}(U)H_{\phi''}(U)L_{\phi}(U)\frac{(C^{2s}_{\omega''}-C^{2s}_{\phi})(C^{2s}_{\phi''}-C^{2s}_{\phi})}
  {(C^{2s}_{\omega''}-C^{2s}_{\omega'})(C^{2s}_{\phi''}-C^{2s}_{\omega'})}\\\nonumber
  -H_{\omega''}(U)H_{\phi''}(U)H_{\omega'}(U)\Bigg](f^{(2)}_{\phi NN}/f_{\phi})\Bigg\}
\end{eqnarray}
with 4 free parameters
$(f^{(2)}_{\phi'NN}/f_{\phi'})$, $(f^{(2)}_{\omega NN}/f_{\omega})$,
$(f^{(2)}_{\phi NN}/f_{\phi}), t^{2s}_{in}$, and
\begin{eqnarray}\label{FN2v}
  F^N_{2v}[X(t)]=\Bigg(\frac{1-X^2}{1-X^2_N}\Bigg)^6\Bigg\{\frac{1}{2}(\mu_p-\mu_n-1)L_\rho(X)L_{\rho'}(X)H_{\rho''}(X)\Bigg\}
\end{eqnarray}
dependent on only 1 free parameter
$t^{2v}_{in}$ with
\begin{eqnarray}
  L_r(V)=\frac{(V_N-V_r)(V_N-V^*_r)(V_N-1/V_r)(V_N-1/V^*_r)}{(V-V_r)(V-V^*_r)(V-1/V_r)(V-1/V^*_r)},\\\label{eq19}
  C^{1s}_r=\frac{(V_N-V_r)(V_N-V^*_r)(V_N-1/V_r)(V_N-1/V^*_r)}{-(V_r-1/V_r)(V_r-1/V^*_r)}, r=\omega, \phi \nonumber
\end{eqnarray}
\begin{eqnarray}
  H_l(V)=\frac{(V_N-V_l)(V_N-V^*_l)(V_N+V_l)(V_N+V^*_l)}{(V-V_l)(V-V^*_l)(V+V_l)(V+V^*_l)},\\\label{eq20}
  C^{1s}_l=\frac{(V_N-V_l)(V_N-V^*_l)(V_N+V_l)(V_N+V^*_l)}{-(V_l-1/V_l)(V_l-1/V^*_l)}, l=
  \omega'', \phi'', \omega', \phi' \nonumber
\end{eqnarray}
\begin{eqnarray}
  L_k(W)=\frac{(W_N-W_k)(W_N-W^*_k)(W_N-1/W_k)(W_N-1/W^*_k)}{(W-W_k)(W-W^*_k)(W-1/W_k)(W-1/W^*_k)},\\ \label{eq21}
  C^{1v}_k=\frac{(W_N-W_k)(W_N-W^*_k)(W_N-1/W_k)(W_N-1/W^*_k)}{-(W_k-1/W_k)(W_k-1/W^*_k)}, k=\rho'',
  \rho', \rho \nonumber
\end{eqnarray}
\begin{eqnarray}
  L_r(U)=\frac{(U_N-U_r)(U_N-U^*_r)(U_N-1/U_r)(U_N-1/U^*_r)}{(U-U_r)(U-U^*_r)(U-1/U_r)(U-1/U^*_r)},\\ \label{eq22}
  C^{2s}_r=\frac{(U_N-U_r)(U_N-U^*_r)(U_N-1/U_r)(U_N-1/U^*_r)}{-(U_r-1/U_r)(U_r-1/U^*_r)}, r=\omega, \phi \nonumber
\end{eqnarray}
\begin{eqnarray}
  H_l(U)=\frac{(U_N-U_l)(U_N-U^*_l)(U_N+U_l)(U_N+U^*_l)}{(U-U_l)(U-U^*_l)(U+U_l)(U+U^*_l)},\\ \label{eq23}
  C^{2s}_l=\frac{(U_N-U_l)(U_N-U^*_l)(U_N+U_l)(U_N+U^*_l)}{-(U_l-1/U_l)(U_l-1/U^*_l)}, l=
  \omega'', \phi'', \omega', \phi' \nonumber
\end{eqnarray}
\begin{eqnarray}
  L_k(X)=\frac{(X_N-X_k)(X_N-X^*_k)(X_N-1/X_k)(X_N-1/X^*_k)}{(X-X_k)(X-X^*_k)(X-1/X_k)(X-1/X^*_k)},\\\label{eq24}
  C^{2v}_k=\frac{(X_N-X_k)(X_N-X^*_k)(X_N-1/X_k)(X_N-1/X^*_k)}{-(X_k-1/X_k)(X_k-1/X^*_k)}, k=\rho', \rho \nonumber
\end{eqnarray}
\begin{eqnarray}
  H_{\rho''}(X)=\frac{(X_N-X_{\rho''})(X_N-X^*_{\rho''})(X_N+X_{\rho''})(X_N+X^*_{\rho''})}
  {(X-X_{\rho''})(X-X^*_{\rho''})(X+X_{\rho''})(X+X^*_{\rho''})},\\ \label{eq25}
  C^{2v}_{\rho''}=\frac{(X_N-X_{\rho''})(X_N-X^*_{\rho''})(X_N+X_{\rho''})(X_N+X^*_{\rho''})}
  {-(X_{\rho''}-1/X_{\rho''})(X_{\rho''}-1/X^*_{\rho''})}.\nonumber
\end{eqnarray}

   Whereas the Dirac $F^p_1(t)$ and Pauli $F^p_2(t)$ proton FFs, as it is seen from the previous, are very
effective for theoretical considerations of the proton EM structure, an extraction of experimental information on it from measured cross sections and polarizations, by the proton electric $G^p_E(t)$ and proton magnetic $G^p_M(t)$ FFs is more comfortable. They appear e.g. in the total cross section of $e^+e^- \to p \bar p$ process  \cite{BPZZ}
\begin{eqnarray}\label{totcspp}
 \sigma_{tot}(e^+e^- \to p \bar p)=\frac{4 \pi \alpha^2 C \beta_p(t)}{3 t}
 [|G^p_M(t)|^2+\frac{2m_p}{t}|G^p_E(s)|^2]
\end{eqnarray}
with $\beta_p(t)=\sqrt{1-\frac{4 m^2_p}{t}}$ and $C$ to be the Coulomb enhancement factor, without interference term, unlike the case if Dirac and Pauli FFs appear in the same total cross section.

   The relations between both sets of FFs are

\begin{eqnarray}\label{pEMFFs}
  G^p_E(t)=[F^N_{1s}(t)+F^N_{1v}(t)]+
  \frac{t}{4 m^2_p}[F^N_{2s}(t)+F^N_{2v}(t)]\\
  G^p_M(t)=[F^N_{1s}(t)+F^N_{1v}(t)]+[F^N_{2s}(t)+F^N_{2v}(t)]\nonumber
\end{eqnarray}
with normalizations
\begin{eqnarray}
  G^p_E(0)= 1;\quad G^p_M(0)=\mu_p;
\end{eqnarray}
and
\begin{eqnarray}
  F^N_{1s}(0)=F^N_{1v}(0)=\frac{1}{2};\quad F^N_{2s}(0)=\frac{1}{2}(\mu_p+\mu_n-1);\quad F^N_{2v}(0)=\frac{1}{2}(\mu_p-\mu_n-1),
\end{eqnarray}
where $\mu_p$ and $\mu_n$ are the magnetic moments of the proton and neutron, respectively.

   The model is in fact a well-matched unification of pole contributions of unstable vector mesons with cut structure in the complex plane $t$,
whereby these cuts represent so-called continua contributions generated by exchange of more than one particle in the corresponding Feynman diagrams and they secure FFs imaginary parts to be different from zero just beyond the lowest possible thresholds on the positive real axis, as it is required by the FFs unitarity conditions.

\begin{figure}
    \includegraphics[width=0.35\textwidth]{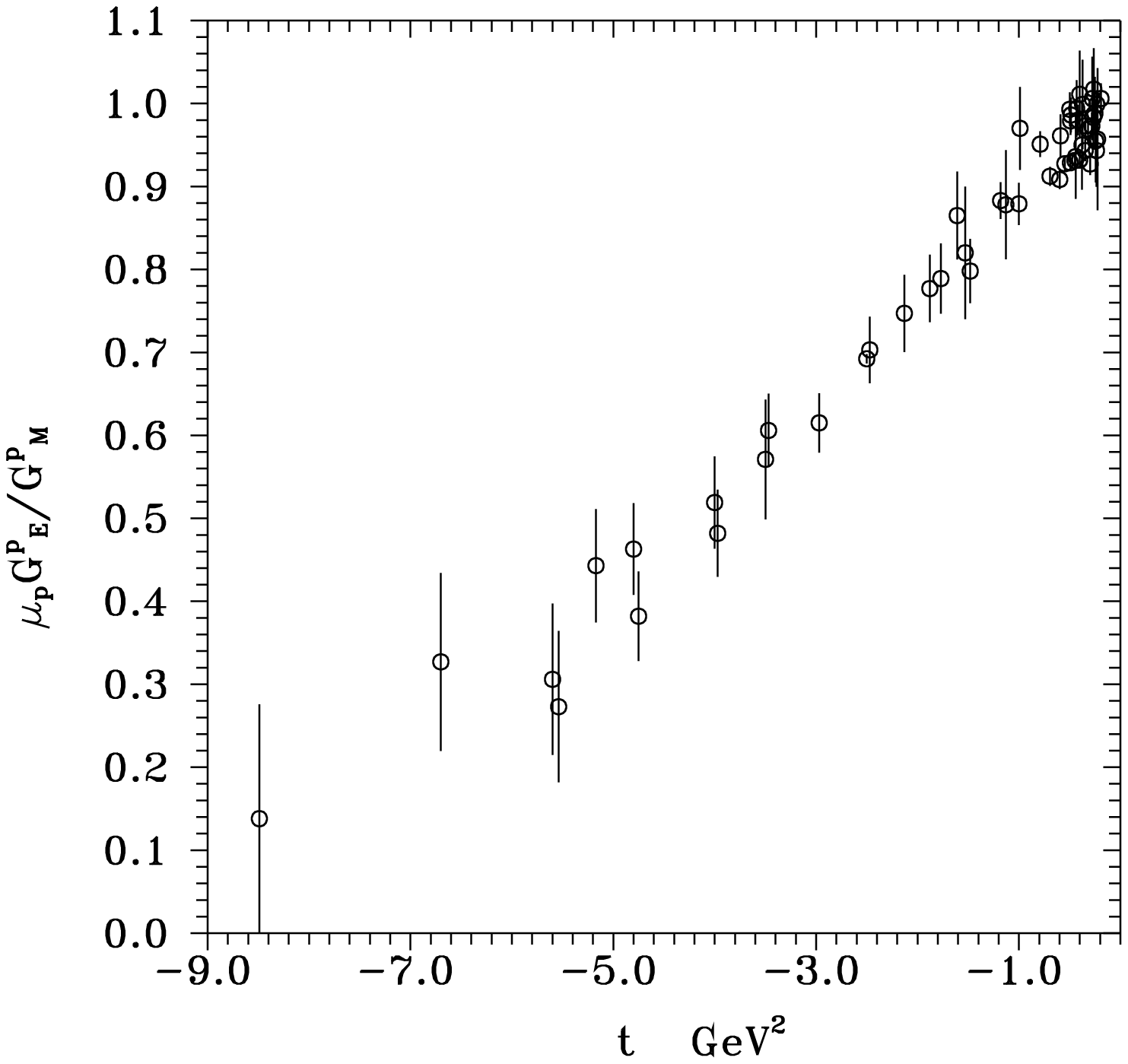}\hspace{0.3cm}
    \includegraphics[width=0.35\textwidth]{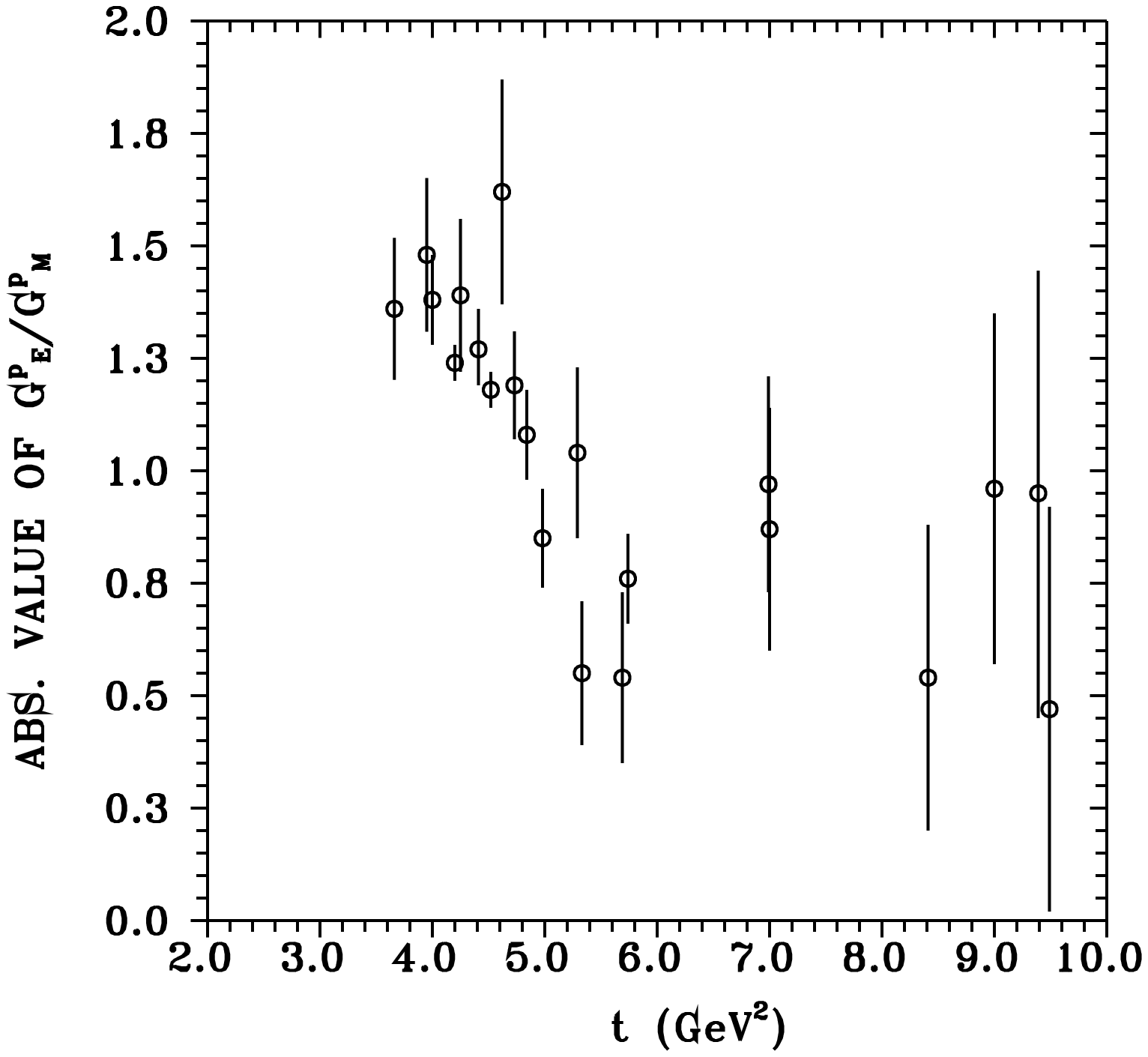}\\
\caption{Experimental data on the ratios of the proton electric to magnetic FFs in space-like and time-like
regions.\label{rpemstd}}
\end{figure}

\begin{figure}
    \includegraphics[width=0.35\textwidth]{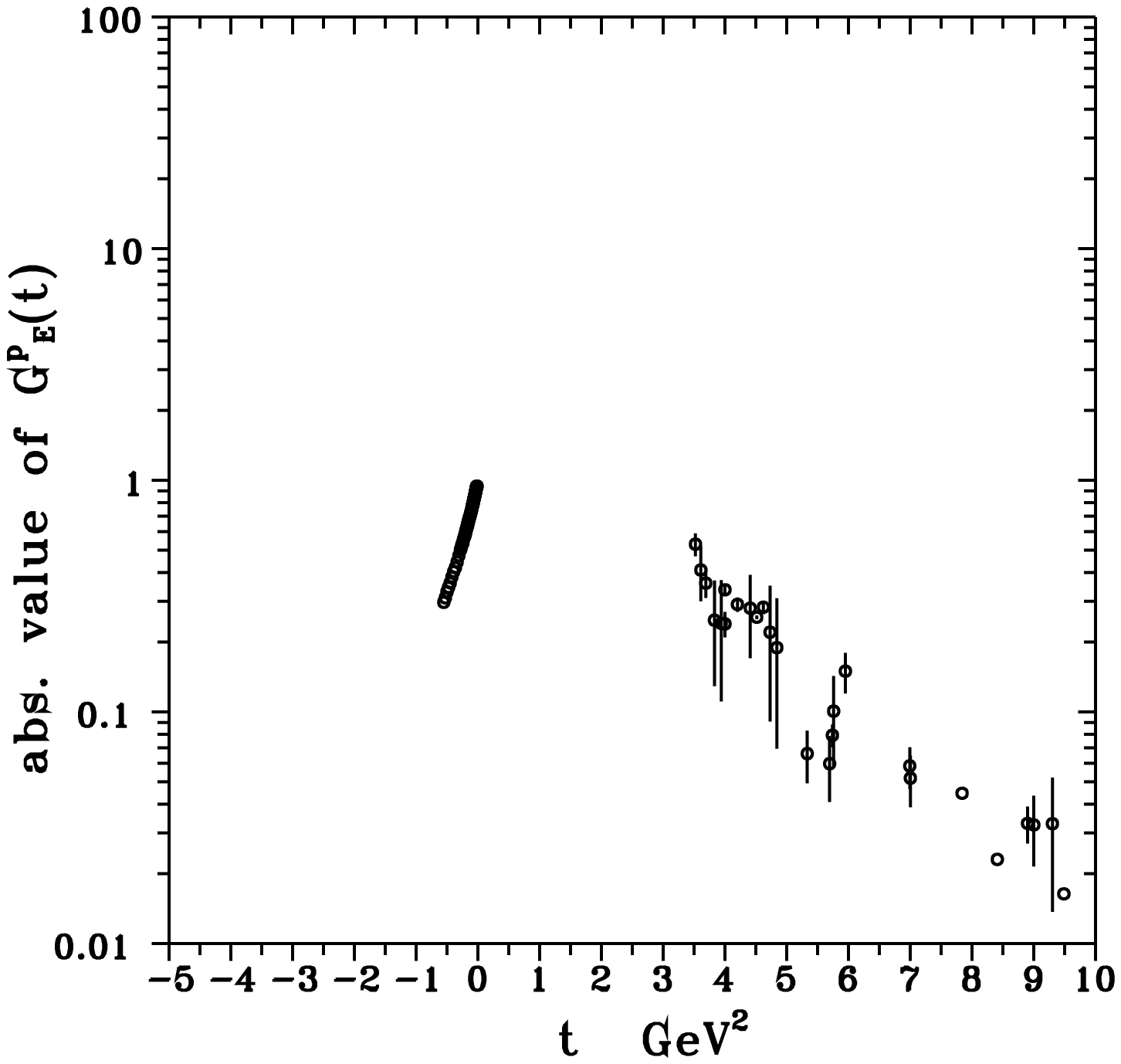}\hspace{0.3cm}
    \includegraphics[width=0.35\textwidth]{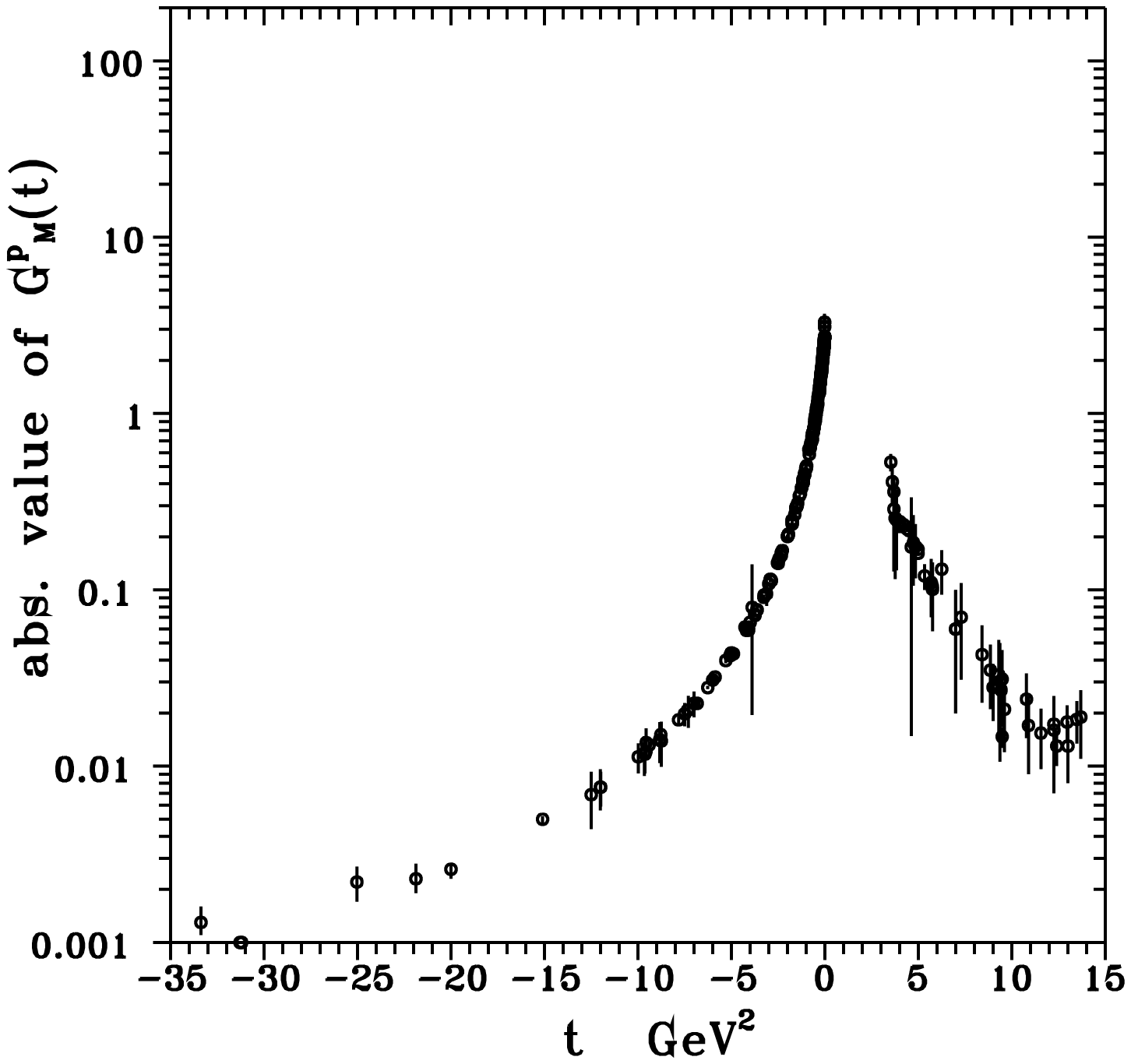}\\
\caption{Experimental data on the proton electric and magnetic FFs in space-like and time-like regions.\label{pemstd}}
\end{figure}

   There are approximately 460 more or less reliable experimental points on the proton EM FFs $G^p_E(t), G^p_M(t)$ \cite{jones}-\cite{Ab1} to be
graphically presented in Figs. \ref{rpemstd}-\ref{pemstd}.

   The results of their simultaneous analysis by the proton EM structure $U\&A$ model given by (\ref{FN1s})-(\ref{FN2v}) and (\ref{pEMFFs}) with 12 free parameters with a clear physical meaning are given in TABLE.

\bigskip

TABLE: Values of free parameters of the proton EM structure $U\&A$ model evaluated in the analysis of the proton EM FFs data. \\
  $\chi^2/ndf=1.74;$ \quad $s^{1s}_{in}= (1.6750\pm 0.0363) GeV^2; s^{1v}_{in}= (2.9683\pm 0.0091) GeV^2;$\\
  $s^{2s}_{in}= (1.8590\pm 0.0023) GeV^2; s^{2v}_{in}= (2.4425\pm 0.0208) GeV^2;$\\
  $(f^{(1)}_{\omega' NN}/f_{\omega'})= -0.2937\pm 0.0015; (f^{(1)}_{\phi' NN}/f_{\phi'})= -0.5298\pm 0.0027;\\
   (f^{(1)}_{\omega NN}/f_{\omega})= 0.6384\pm 0.0025; (f^{(1)}_{\phi NN}/f_{\phi})= -0.0271\pm 0.0005;\\
   (f^{(2)}_{\phi' NN}/f_{\phi'})= 0.3075\pm 0.0156; (f^{(2)}_{\omega NN}/f_{\omega})= 0.1676\pm 0.0377;\\
   (f^{(2)}_{\phi NN}/f_{\phi})= 0.1226\pm 0.0035; (f^{(1)}_{\rho NN}/f_{\rho})= -0.0802\pm 0.0014;$\\

\begin{figure}
    \includegraphics[width=0.35\textwidth]{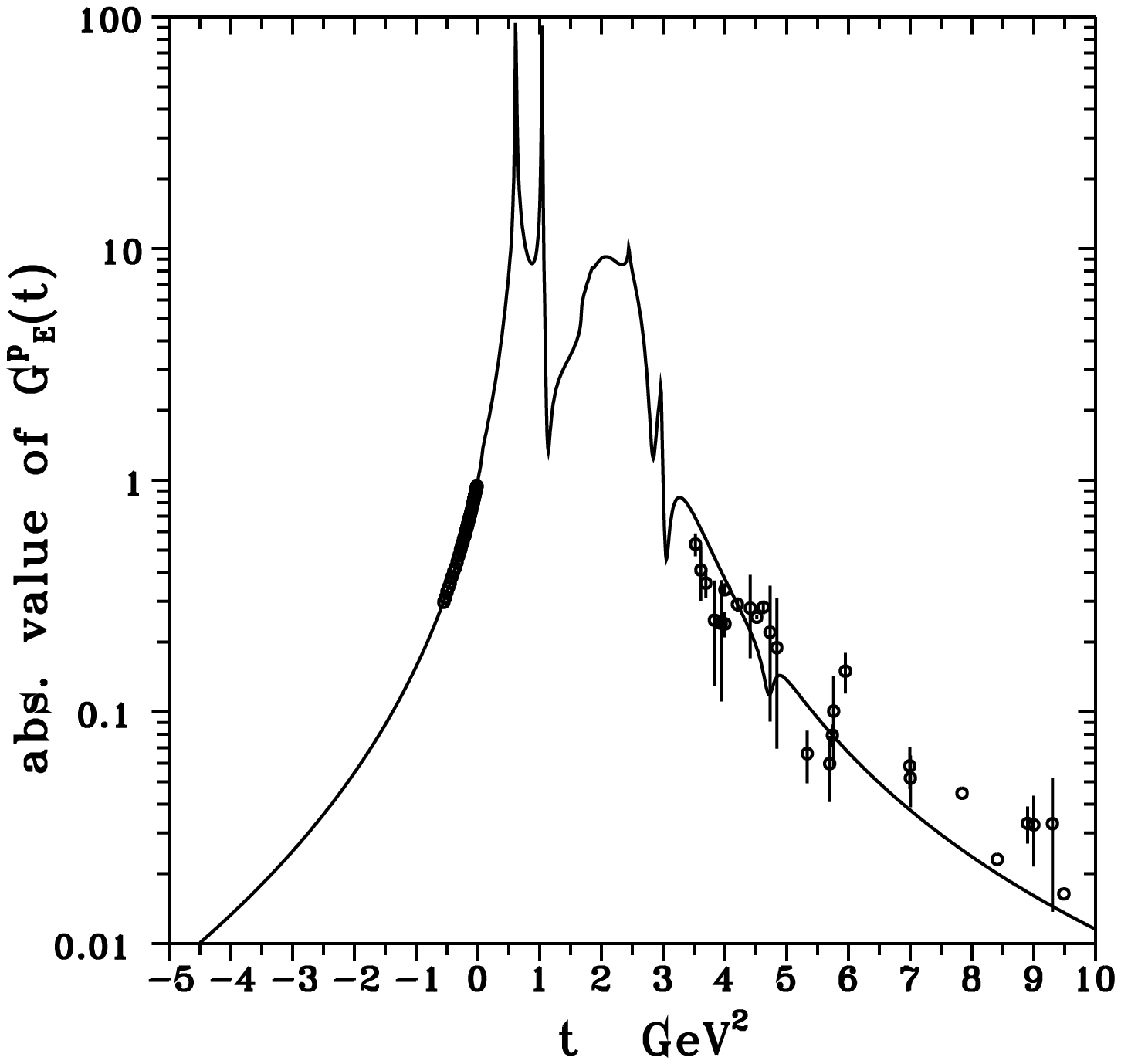}\hspace{0.3cm}
    \includegraphics[width=0.35\textwidth]{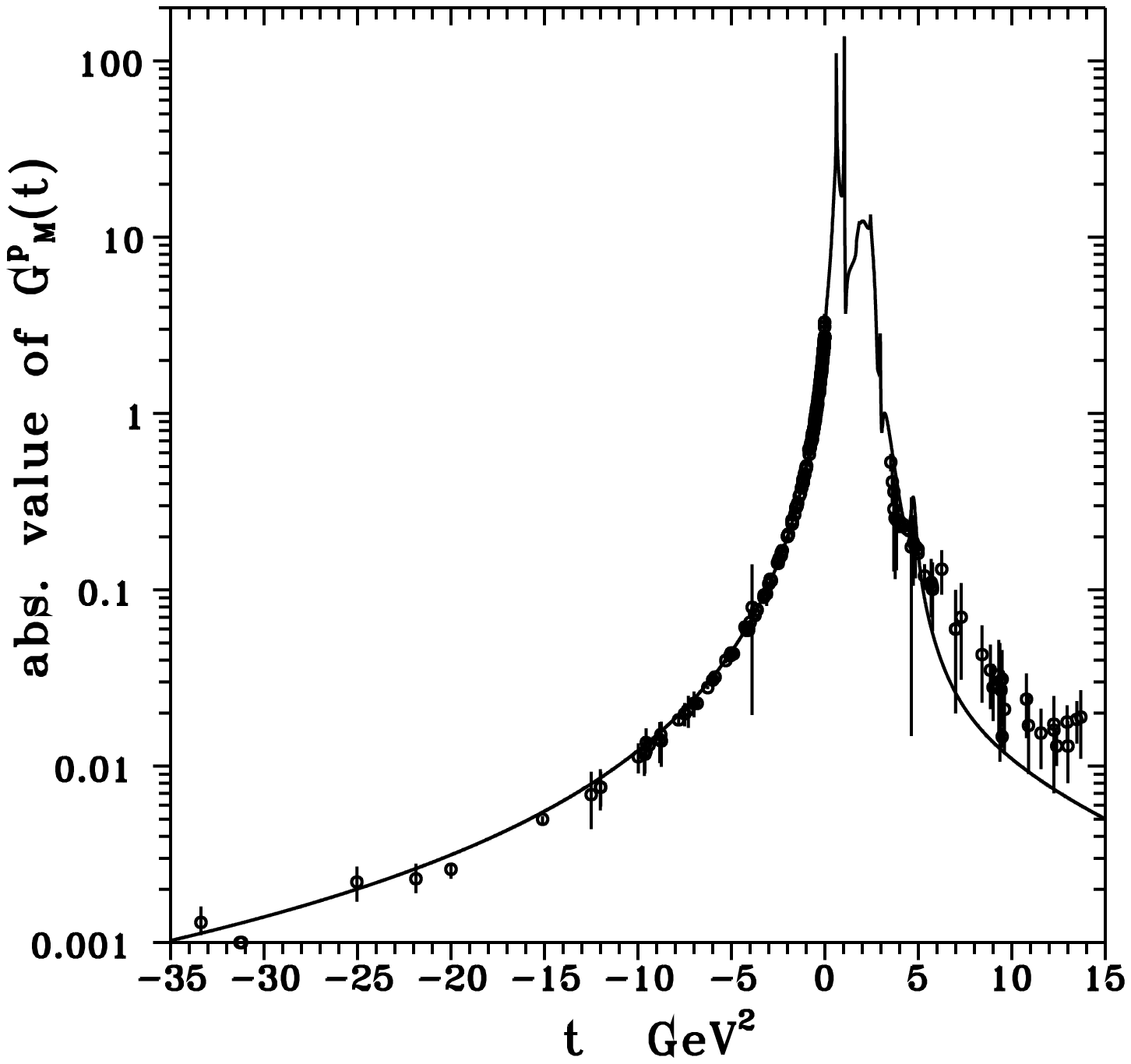}\\
\caption{The proton EM structure $U\&A$ model given by (\ref{FN1s})-(\ref{FN2v}) and (\ref{pEMFFs}) with parameters of TABLE reproduce
existing data quite well.\label{pemstth}}
\end{figure}

   The corresponding behaviors of absolute values of the proton electric and magnetic FFs on the base of these results are given by full lines in Fig. \ref{pemstth},
where also their comparison with exiting data is presented.

\section{Theoretical predictions of neutron electromagnetic form factors behaviors}

   Similarly to the proton one can define the neutron Dirac $ F^n_1(t)$ and Pauli $ F^n_2(t)$ FFs by means of the neutron matrix element of the
EM current $J^{EM}_\mu (0)$
\begin{small}
\begin{eqnarray}
  <n|J^{EM}_\mu (0)|n>=e \bar u(p')\{\gamma_\mu F^n_1(t)+\frac{i}{2m_n}
  \sigma_{\mu \nu}(p'-p)_\mu F^n_2(t)\} u(p),
\end{eqnarray}
\end{small}
with $m_n$ to be the neutron mass. Again mixed transformation properties of the EM current
$J^{EM}_\mu (0)$ under the rotation in the isospin space is utilized, which leads to a splitting of the neutron Dirac $F^n_1(t)$ and Pauli $F^n_2(t)$ FFs to the same flavour-independent isoscalar $F^N_{1s}(t), F^N_{2s}(t)$ and isovector $F^N_{1v}(t), F^N_{2v}(t)$ parts like in the case of the proton as follows
\begin{small}
\begin{eqnarray}
 F^n_1(t)=[F^N_{1s}(t)-F^N_{1v}(t)]\nonumber\\
 F^n_2(t)=[F^N_{2s}(t)-F^N_{2v}(t)]\\\nonumber
\end{eqnarray}
\end{small}
however, now with the minus sign between them as the sign of the third component of the neutron isospin is -1/2.

\begin{figure}
    \includegraphics[width=0.35\textwidth]{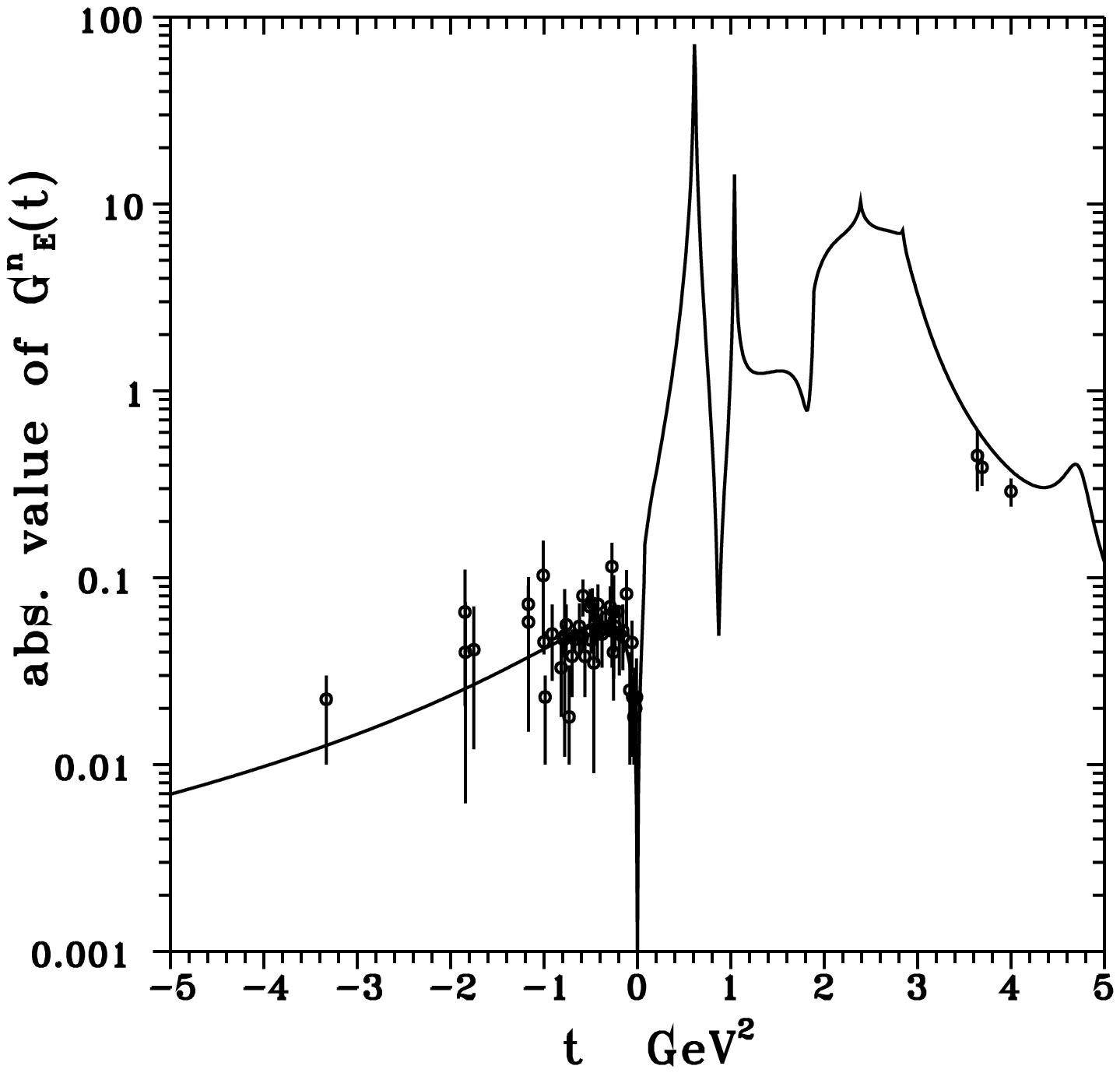}\hspace{0.3cm}
    \includegraphics[width=0.35\textwidth]{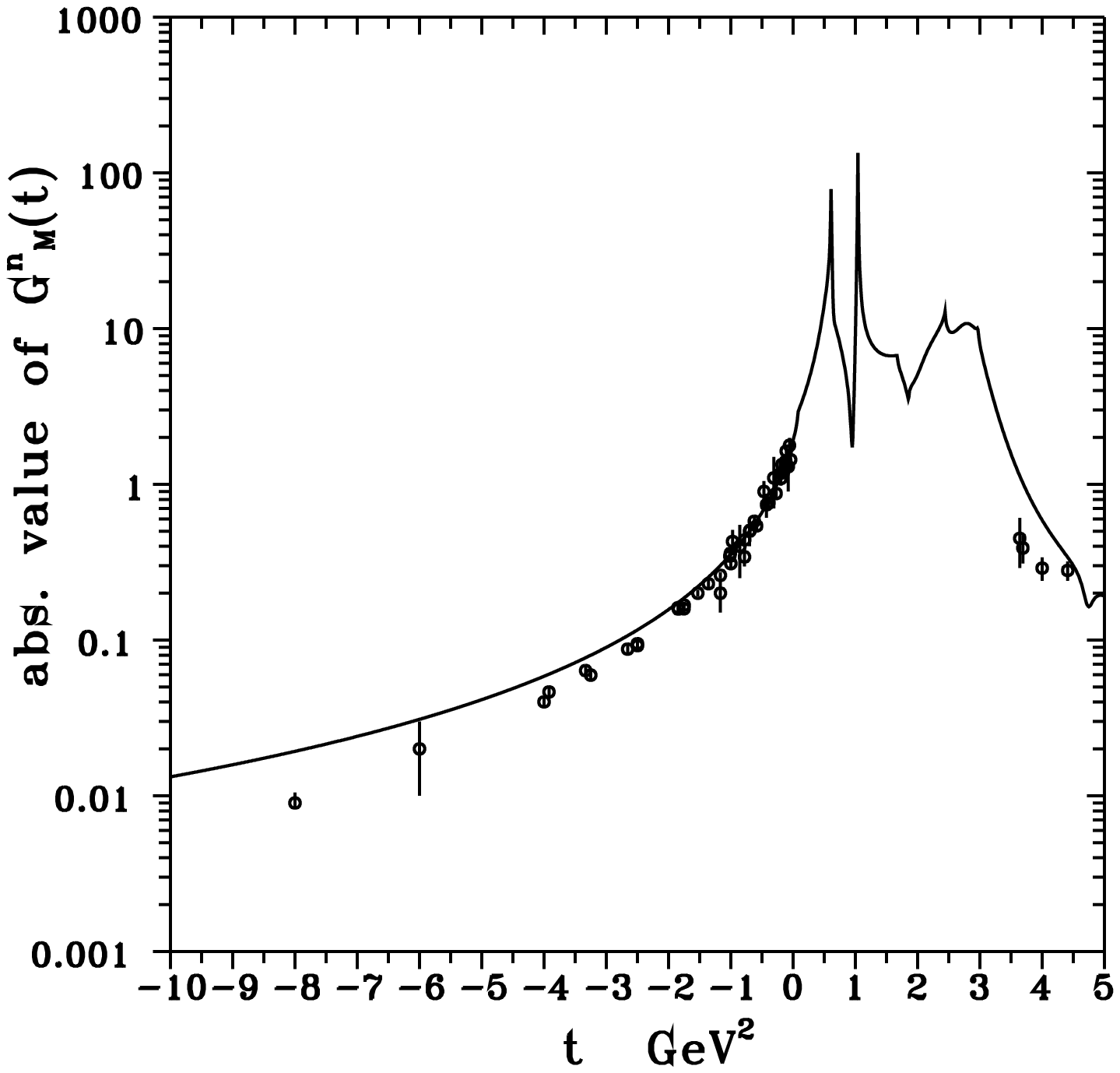}\\
\caption{Predicted absolute value of the neutron electric and magnetic FFs behaviors without the use of any experimental point \cite{hanson}-\cite{riordan} seen in this figures.}\label{nemstth}
\end{figure}

   Then the relations between neutron EM FFs and the flavour-independent neutron Dirac and  Pauli FFs are
\begin{eqnarray}\label{nEMFFs}
  G^n_E(t)=[F^N_{1s}(t)-F^N_{1v}(t)]+
  \frac{t}{4 m^2_n}[F^N_{2s}(t)-F^N_{2v}(t)]\\
  G^n_M(t)=[F^N_{1s}(t)-F^N_{1v}(t)]+[F^N_{2s}(t)-F^N_{2v}(t)]\nonumber
\end{eqnarray}
with normalizations
\begin{eqnarray}
  G^n_E(0)= 0;\quad G^n_M(0)=\mu_n.
\end{eqnarray}

   Since the FFs $F^N_{1s}(t), F^N_{2s}(t)$ and $F^N_{1v}(t), F^N_{2v}(t)$ are completely known by the expressions (\ref{FN1s})-(\ref{FN2v}) with
free parameters from TABLE to be evaluated in the analysis of only the proton EM FFs data, in combination with relations (\ref{nEMFFs}) they
represent the advanced neutron electromagnetic structure $U\&A$ model, which enables to predict behaviors of the absolute values of the neutron electric $G^n_E(t)$ and magnetic $G^n_M(t)$ FFs theoretically, as they are presented by the full lines in Fig. \ref{nemstth}, without the use of any experimental point \cite{hanson}-\cite{riordan} on the neutron EM FFs data also shown there for a comparison. The results are encouraging.

\section{Conclusions}

   The absolute values of the neutron EM FFs, $G^n_E(t), G^n_M(t)$, are predicted theoretically in the whole region of their definition
without exploiting of any experimental point on them. This could be carried out by the fact that the proton EM structure $U\&A$ model given by (\ref{FN1s})-(\ref{FN2v}) and (\ref{pEMFFs}) and the neutron EM structure $U\&A$ model given by (\ref{FN1s})-(\ref{FN2v}) and (\ref{nEMFFs}) are compound of the same flavour-independent isoscalar $F^N_{1s}(t), F^N_{2s}(t)$ and isovector $F^N_{1v}(t), F^N_{2v}(t)$ parts of the corresponding Dirac and Pauli FFs, and it does not matter that in the case of the proton with "+" sign and in the case of the neutron with "-" sign, however, they depend on the same free parameters.

   So, these parameters can be numerically evaluated either in the analysis of the neutron EM FFs data by means of the neutron EM structure $U\&A$ model and then one
predicts absolute values of the proton EM FFs behaviors, or in the analysis of the proton EM FFs data by means of the proton EM structure $U\&A$ model and then one predicts absolute values of the neutron EM FFs behaviors. Naturally, as at present the proton EM FFs are more reliable and more precise, we have preferred the second case.

   On the conclusion we would like to emphasize that if more precise and more reliable proton EM FFs data will appear in as wide region of their definition as
possible, the more reliable predictions for the absolute values of the neutron EM FFs will be achieved by the approach discussed in this paper.

\medskip

   The authors would like to thank Eric Bartos for valuable discussions.
   The support of the Slovak Grant Agency for Sciences VEGA, grant No.2/0153/17, is acknowledged.

\end{document}